\documentclass[11pt]{elsarticle}

\usepackage{lineno,hyperref}
\modulolinenumbers[5]

\journal{Journal of Computational Physics}

\usepackage[margin=2.5cm]{geometry}

\usepackage{natbib}
\usepackage{bm}
\usepackage{amsmath}
\usepackage{amsfonts}
\usepackage{amssymb}
\usepackage{amsthm}
\usepackage{mathrsfs}
\usepackage{mathtools}
\usepackage{empheq}
\usepackage{cleveref}
\usepackage{float}

\usepackage[monochrome,dvipsnames]{xcolor}

\usepackage{booktabs}

\renewcommand{\vec}[1]{\mathbf{#1}}

\usepackage{xspace}
\let\storeBeta=\beta
\renewcommand\beta{\relax\ifmmode{\storeBeta}\else{$\storeBeta$}\fi\xspace}

\let\storeAlpha=\alpha
\renewcommand\alpha{\relax\ifmmode{\storeAlpha}\else{$\storeAlpha$}\fi\xspace}

\newcommand{\pd}[2]{\frac{\partial #1}{\partial #2}} 
\newcommand{\pdd}[2]{\frac{\partial^2 #1}{\partial #2^2}} 

\newcommand{\norm}[1]{\left\lVert#1\right\rVert}

\hypersetup{
	colorlinks,
	linkcolor={blue!50!black},
	citecolor={blue!50!black},
	urlcolor={blue!80!black},
	anchorcolor = {blue!80!black},
	filecolor = {blue!80!black},
	menucolor = {blue!80!black},
	runcolor = {blue!80!black}
}

\usepackage{tgpagella}
\usepackage{newpxmath}

\usepackage{siunitx}
\usepackage{tikz} 
\usetikzlibrary{backgrounds}
\usetikzlibrary{shapes,arrows,positioning,intersections,quotes}
\usepackage{tcolorbox}
\usepackage{pgfplots}

\pgfplotsset{compat=newest,scaled y ticks=true} 
\usepgfplotslibrary{units} 
\usepgfplotslibrary{colorbrewer}
\usepackage{pgfplotstable}

\newcommand{\logLogSlopeTriangle}[5]
{
	
	\pgfplotsextra
	{
		\pgfkeysgetvalue{/pgfplots/xmin}{\xmin}
		\pgfkeysgetvalue{/pgfplots/xmax}{\xmax}
		\pgfkeysgetvalue{/pgfplots/ymin}{\ymin}
		\pgfkeysgetvalue{/pgfplots/ymax}{\ymax}
		
		\pgfmathsetmacro{\xArel}{#1}
		\pgfmathsetmacro{\yArel}{#3}
		\pgfmathsetmacro{\xBrel}{#1-#2}
		\pgfmathsetmacro{\yBrel}{\yArel}
		\pgfmathsetmacro{\xCrel}{\xArel}
		
		\pgfmathsetmacro{\lnxB}{\xmin*(1-(#1-#2))+\xmax*(#1-#2)} 
		\pgfmathsetmacro{\lnxA}{\xmin*(1-#1)+\xmax*#1} 
		\pgfmathsetmacro{\lnyA}{\ymin*(1-#3)+\ymax*#3} 
		\pgfmathsetmacro{\lnyC}{\lnyA+#4*(\lnxA-\lnxB)}
		\pgfmathsetmacro{\yCrel}{\lnyC-\ymin)/(\ymax-\ymin)} 
		
		\coordinate (A) at (rel axis cs:\xArel,\yArel);
		\coordinate (B) at (rel axis cs:\xBrel,\yBrel);
		\coordinate (C) at (rel axis cs:\xCrel,\yCrel);
		
		\draw[#5]   (A)-- node[pos=0.5,anchor=north] {\tiny1}
		(B)-- 
		(C)-- node[pos=0.5,anchor=west] {\tiny#4}
		cycle;
	}
}

\usepackage[colorinlistoftodos,prependcaption,textsize=normalsize,disable]{todonotes}

\begin{document}

\begin{frontmatter}

\title{Assessment of an energy-based surface tension model for simulation of two-phase flows using second-order phase field methods}

\author[myaddress]{Shahab Mirjalili\corref{mycorrespondingauthor}}
\ead{ssmirjal@stanford.edu}
\cortext[mycorrespondingauthor]{Corresponding author}
\author[myaddress]{Makrand A. Khanwale}
\ead{khanwale@stanford.edu}
\author[myaddress]{Ali Mani}
\ead{alimani@stanford.edu}
\address[myaddress]{Center for Turbulence Research, Stanford University, Stanford, CA 94305, USA}

\begin{abstract}
Second-order phase field models have emerged as an attractive option for capturing the advection of interfaces in two-phase flows. Prior to these, state-of-the-art models based on the Cahn-Hilliard equation, which is a fourth-order equation, allowed for the derivation of surface tension models through thermodynamic arguments. 
In contrast, the second-order phase field models do not follow a known energy law, and deriving a surface tension term for these models using thermodynamic arguments is not straightforward.
In this work, we justify that the energy-based surface tension model from the Cahn-Hilliard context can be adopted for second-order phase field models as well and assess its performance. We test the surface tension model on three different second-order phase field equations; the conservative diffuse interface model of~\citet{Chiu_and_Lin}, and two models based on the modified Allen-Cahn equation introduced by~\citet{Sun2007}.
 \textcolor{ForestGreen}{
Additionally, we draw the connection between the energy-based model with a localized variation of the continuum surface force (CSF) model.
Using canonical tests, we illustrate the lower magnitude of spurious currents, better accuracy, and superior convergence properties of the energy-based surface tension model compared to the CSF model, which is a popular choice used in conjunction with second-order phase field methods, and the localized CSF model.} Importantly, in terms of computational expense and parallel efficiency, the energy-based model incurs no penalty compared to the CSF models. 
\end{abstract}

\begin{keyword}
two-phase flow, phase field, surface tension, diffuse interface, continuous surface force, Allen-Cahn
\end{keyword}

\end{frontmatter}

\section{Introduction}
\label{sec:introduction}
In recent years, there has been considerable interest in second-order phase field equations that are suitable for modeling two-phase flows \citep{Sun2007, Chiu_and_Lin, Mirjalili_boundedness, Huang2020}. Unlike models based on the Cahn-Hilliard equation \citep{Jacqmin1999, Abels2012}, these equations are not derived from an underlying thermodynamic basis, and as a result, it is difficult to derive energy-based surface tension terms for these models. Based on the second law of thermodynamics, the Korteweg stress tensor has been used successfully to model surface tension effects in the context of Cahn-Hilliard models~\citep{Abels2012, Khanwale2020}. In this work, starting from the Korteweg stress tensor, we derive an energy-based surface tension force model for second-order phase field equations. Using canonical test problems on three different phase field equations, we show that the energy-based model has better properties than the continuum surface model (CSF), which is the highly explored alternative~\citep{Brackbill1992, Chiu_and_Lin, Mirjalili2021} and a localized variation of CSF. Specifically, the proposed model (1) is accurate and generates low spurious currents, (2) requires lower regularity compared to the direct implementation of the Korteweg stress tensor, (3) does not require computation of curvature, (4) is amenable to a wide range of spatio-temporal discretization schemes. \textcolor{ForestGreen}{We note regarding the third item that while the proposed surface tension model does not need curvature computation, some second-order phase field equations explicitly involve curvature fields.}\todo[color=ForestGreen]{R2: \#2.1}

\section{Model derivation}
\label{sec:model_derivation}

To derive the energy-based surface tension model, let us consider a general phase field equation for incompressible flows in the form given by
\begin{equation}
	\pd{\phi}{t} + \pd{}{x_j}\left(u_j \phi\right) = g(\phi, \mathbf{x}).
\end{equation}
The role of $g(\phi, \mathbf{x})$ is to regularize the interface. Traditional phase field models derive this right-hand side term such that it minimizes a free-energy functional. Most commonly, the Ginzburg-Landau free energy is utilized, corresponding to the chemical potential $\mu$ ~\citep{Gurtin1996, Abels2012} given by 
\begin{equation}
	\mu = \frac{\hat{\sigma}}{\epsilon}\psi^\prime - \hat{\sigma}\epsilon\pd{}{x_j}\left({\pd{\phi}{x_j}}\right),\;\;\text{where}\;\;\psi =  \frac{1}{2}\left[\phi(1-\phi)\right]^2\;\;\text{and}\;\;\psi^\prime =  \phi\left(1-\phi\right)\left(1-2\phi\right).
	\label{eqn:ac_mu}
\end{equation}
$\epsilon$ is the interfacial thickness parameter, and $\psi$ is the free energy for a phase field variable, $\phi\in [0,1]$. Additionally, $\hat{\sigma}$ is a constant related to surface energy, given by $\hat{\sigma}=6\sigma$\footnote{Note that for Cahn-Hilliard models, instead of the factor 6 we have the factor of $3/2\sqrt{2}\approx 1$. This distinction is due to the difference of the equilibrium profile of $\phi$ in the Cahn-Hilliard based models given by $\tanh{\left(s/\sqrt{2}\epsilon\right)}$ and $\phi \in [-1, 1]$, in contrast to \textcolor{cyan}{$[1 + \tanh{\left(s/2\epsilon\right)}]/2$} and $\phi \in [0, 1]$ for phase field models considered here ($s$ is the signed distance from the interface).}, where $\sigma$ is the surface tension coefficient.  In the special case of the Allen-Cahn model we would have $g(\phi, \mathbf{x}) = -\mu$. Note that the Allen-Cahn equation admits an energy dissipation law similar to the Cahn-Hilliard equation. However, the lack of mass conservation, in addition to the curvature-driven flow it generates, renders it unsuitable for simulation of two-phase flows, motivating the introduction of the alternative second-order phase field models we consider herein~\citep{Sun2007, Chiu_and_Lin, Mirjalili2021}.  

Historically, models based on the Cahn-Hilliard equation, where $g(\phi, \mathbf{x}) = \nabla^2\mu$, have been used for the simulation of two-phase flows. In this context, one can derive a surface tension model through the second law of thermodynamics. The surface tension force in this model is proportional to the divergence of the so-called Korteweg stress tensor,  $\pd{}{x_j}\left({\pd{\phi}{x_i}\pd{\phi}{x_j}}\right)$, which prescribes an energy dissipation law ~\citep{Shen2010,Abels2012,Guo2017,Khanwale2020}. \textcolor{ForestGreen}{While  this modeling choice for surface tension does not preserve total energy, it imparts stability and robustness to the Cahn-Hilliard-Navier-Stokes system\citep{Jacqmin1999}.\todo[color=ForestGreen]{R2: \#2.2}} 
However, the more recently introduced second-order phase field equations for simulating interfacial flows do not admit an energy dissipation law as they are not gradient flows to any known energy functional. As such, these equations have typically been coupled to the momentum transport equation via the CSF model, $F_i=\sigma\kappa\pd{\phi}{x_i}$, where $\kappa$ is the curvature ~\citep{Chiu_and_Lin,Mirjalili2021}, defined as
\begin{equation}
    \kappa=-\pd{}{x_j}\left(\frac{\pd{\phi}{x_j}}{\left(\pd{\phi}{x_k}\pd{\phi}{x_k}\right)^{1/2}}\right).
	\label{eqn:curvature}
\end{equation}

In this work, we return to the Korteweg stress tensor to assess its suitability for second-order phase field equations. Due to better regularity, the flexibility offered in numerical discretization, and the well-balancedness of volumetric surface tension models compared to integral formulations ~\citep{Popinet2018}, we utilize the equivalent ~\citep{Kim2005} surface tension force model,
	\begin{equation}
	    {F}_{i}=
	    \mu\frac{\partial\phi}{\partial{x}_{i}}.
	    \label{eqn:eb_st}
	\end{equation}

While this surface tension force model has been used and derived in the literature for Cahn-Hilliard models~\citep{Jacqmin1999, Kim2005, Abels2012, Khanwale2020}, for completeness of narrative, we reproduce its derivation here. Starting from the surface tension force based on the Korteweg stress tensor and using the product rule:
\begin{equation}
	\begin{split}
		\hat{\sigma} \epsilon \pd{}{x_j}\left({\pd{\phi}{x_i}\pd{\phi}{x_j}}\right) &= 
		\hat{\sigma} \epsilon \pd{\phi}{x_i}\left(\pd{}{x_j}\left({\pd{\phi}{x_j}}\right)\right) 
		+ \hat{\sigma} \epsilon \pd{\phi}{x_j}\left(\pd{}{x_j}\left({\pd{\phi}{x_i}}\right)\right) \\
		& =
		\hat{\sigma} \epsilon \pd{\phi}{x_i}\left(\pd{}{x_j}\left({\pd{\phi}{x_j}}\right)\right) 
		+\hat{\sigma} \epsilon \frac{1}{2}\pd{}{x_i}\left(  \pd{\phi}{x_j} \pd{\phi}{x_j} \right).
		\label{eqn:forcing_expan}
	\end{split}
\end{equation}
We manipulate this expression to write it in terms of the chemical potential:
\begin{equation}
	\begin{split}
		\hat{\sigma} \epsilon \pd{}{x_j}\left({\pd{\phi}{x_i}\pd{\phi}{x_j}}\right) &= \hat{\sigma} \epsilon \pd{\phi}{x_i}\left(\pd{}{x_j}\left({\pd{\phi}{x_j}}\right)\right) 
		+\hat{\sigma} \epsilon \frac{1}{2}\pd{}{x_i}\left(  \pd{\phi}{x_j} \pd{\phi}{x_j} \right)
		+ \hat{\sigma} \epsilon \frac{\psi^\prime}{\epsilon^2}\, \pd{\phi}{x_i} 
		- \hat{\sigma} \epsilon \frac{\psi^\prime}{\epsilon^2} \, \pd{\phi}{x_i},\\ 
		&=   \pd{\phi}{x_i}\left(\frac{\hat{\sigma}}{\epsilon}\epsilon^2\pd{}{x_j}\left({\pd{\phi}{x_j}}\right) -  \frac{\hat{\sigma}}{\epsilon}\psi^\prime \, \right)  
		+\hat{\sigma} \epsilon\frac{1}{2}\pd{}{x_i}\left(  \pd{\phi}{x_j} \pd{\phi}{x_j} \right)
		+ \hat{\sigma} \epsilon\frac{\psi^\prime}{\epsilon^2}  \, \pd{\phi}{x_i}.
		\label{eqn:forcing_expan_f}
	\end{split}
\end{equation}
	
The expression in the parenthesis in the first term can be replaced using the definition of $\mu$ \cref{eqn:ac_mu}, which leads to:
\begin{equation}
	\begin{split}
		\hat{\sigma} \epsilon\pd{}{x_j}\left({\pd{\phi}{x_i}\pd{\phi}{x_j}}\right) &= 
		-\mu\pd{\phi}{x_i}   
		+\hat{\sigma} \epsilon\frac{1}{2}\pd{}{x_i}\left(  \pd{\phi}{x_j} \pd{\phi}{x_j} \right)
		+ \hat{\sigma} \epsilon\frac{\psi^\prime}{\epsilon^2}  \, \pd{\phi}{x_i},\\
		&=
		-\mu\pd{\phi}{x_i}   
		+\pd{}{x_i}\left( \hat{\sigma} \epsilon \frac{1}{2}\pd{\phi}{x_j} \pd{\phi}{x_j} 
		+ \hat{\sigma} \frac{\psi}{\epsilon}  \,\right).
	\end{split} 
	\label{eqn:forcing_expan_mu}
\end{equation}

Recognize that the last term in \cref{eqn:forcing_expan_mu} can absorb a redefined pressure.  Therefore, the first term in \cref{eqn:forcing_expan_mu} can be used as a surface tension force in the momentum equations. 
	Note that the sign of the surface tension force in \cref{eqn:forcing_expan_mu} is the opposite of the Korteweg stress tensor. Additionally, the sign of $\mu\pd{\phi}{x_i} $ would be positive on the right-hand side of the momentum equation. \textcolor{cyan}{It is worth noting that by drawing analogies with Cahn-Hilliard models, a few prior studies have employed ~\cref{eqn:eb_st} in conjunction with second-order phase field methods\citep{Haghani2018,Huang2020}. This work justifies why this approach is indeed accurate and advantageous compared to using alternative surface tension models.}\todo[color=cyan]{R1: \#1.1}

\subsection{Considered second-order phase field models}
\label{sec:pf_models}
To ensure the generality of our conclusions, we examine the performance of the above energy-based surface tension force on three different second-order phase field equations. We consider (1) the conservative diffuse interface model, denoted by CDI \citep{Chiu_and_Lin, Mirjalili_boundedness}, (2) the modified Allen-Cahn model of \citet{Sun2007}, denoted by SB, and (3) an implicit-capable second-order diffuse interface model we denote by ISODI.  These equations are defined in \cref{tab:models}.

\begin{table}[H]
	\centering
	\centering\footnotesize\setlength\tabcolsep{4pt}
	\begin{tabular}{>{\centering\arraybackslash}m{0.2\linewidth}|>{\centering\arraybackslash}m{0.6\linewidth}|>{\centering\arraybackslash}m{0.2\linewidth}}
		\toprule
		\toprule
		Name & $g\left(\phi\right)$ & Description \\
		\midrule
		\midrule
		CDI 			& 
        $\gamma\pd{}{x_j} \left[\epsilon\pd{\phi}{x_j} - \phi\left(1-\phi\right)n_j\right]$
        &  \textit{Conservative Diffuse Interface} (CDI) model~\citep{Chiu_and_Lin, Mirjalili_boundedness}\\
		\midrule
		\citet{Sun2007} & 
        $\gamma\left[ \epsilon\pdd{\phi}{x_j} - \frac{\phi\left(1 - \phi\right)\left(1 - 2\phi\right)}{\epsilon} + \epsilon\left(\pd{\phi}{x_k}\pd{\phi}{x_k}\right)^{1/2} \kappa \right]$
        & Curvature subtracted Allen-Cahn based model proposed in~\citet{Sun2007}\\
		\midrule		
		ISODI 			& 
        $\gamma\left[ \epsilon\pdd{\phi}{x_j} - \frac{\phi\left(1 - \phi\right)\left(1 - 2\phi\right)}{\epsilon} + \phi(1-\phi) \kappa\right]$
        &  \textit{Implicit capable Second Order Diffuse Interface} (ISODI) is a modification of model in ~\citet{Sun2007} for full time-implicit schemes\\
		\bottomrule
		\bottomrule
	\end{tabular}
    \caption{\textcolor{ForestGreen}{Table describing the second order diffuse interface models under assessment. Note that the interface normal is $n_j = \left.{\pd{\phi}{x_j}}\middle/ {\left(\pd{\phi}{x_k}\pd{\phi}{x_k}\right)^{1/2}}\right.$.} \todo[color=ForestGreen, inline]{R2: \#2.3}
    }
	\label{tab:models}
\end{table}

Note that the ISODI model is derived by modifying the third term in the model by~\citet{Sun2007} \textcolor{ForestGreen}{by invoking the equilibrium assumption}, $\left({\partial\phi}/{\partial x_k}\;\;{\partial\phi}/{\partial x_k}\right)^{1/2}={\phi\left(1-\phi\right)}/{\epsilon}$.  This allows for an easier setup of Newton iteration for fully-implicit schemes. \textcolor{ForestGreen}{It is worth highlighting that since the interface is not necessarily at equilibrium in dynamic cases involving strong deformations, ISODI and the phase field model of \citet{Sun2007} can display different dynamics, which is why we examine both models here.}\todo[color=ForestGreen]{R2: \#2.4}

\subsection{Relation to a localized continuum surface force model}
\label{sec:equivalence_regularized_CSF}

Before implementing the model proposed in~\cref{eqn:eb_st}, we present further analysis of its right-hand side to generate insight regarding its connection and contrast compared to CSF. For second-order phase field equations considered in this work, at equilibrium, the energy-based surface tension model (\cref{eqn:eb_st}) is analytically equivalent to a modified CSF model given by $F_i=\hat{\sigma}\kappa\phi(1-\phi)\pd{\phi}{x_i}$. To prove this, note that at equilibrium, all three models in~\cref{tab:models} yield a hyperbolic tangent profile in the direction normal to the interface, whereby
\begin{equation}
	\left(\pd{\phi}{x_k}\pd{\phi}{x_k}\right)^{1/2}=\frac{\phi\left(1-\phi\right)}{\epsilon}.
	\label{eqn:tanh_identity}
\end{equation}

\textcolor{ForestGreen}{Note that we invoke this assumption just to establish a relationship between the energy-based model and the CSF formulations which are commonly used in the literature. 
The requirement for~\cref{eqn:tanh_identity} to hold is that the interface should be in equilibrium, which is not necessarily true for dynamic systems.\todo[color=ForestGreen]{R2: \#2.5}}

Also, note that, at equilibrium ($g(\phi) = 0$), for the three models presented in~\cref{tab:models}, using~\cref{eqn:tanh_identity}, we have the following equation\footnote{For CDI, here we expand the $g(\phi)$ to get $\epsilon\pdd{\phi}{x_j} - \left(\pd{\phi}{x_k}\pd{\phi}{x_k}\right)^{1/2}\frac{\left(1 - 2\phi\right)}{\epsilon} + \phi\left(1-\phi\right) \kappa$},
\begin{equation}
	\epsilon\pdd{\phi}{x_j} - \frac{\phi\left(1 - \phi\right)\left(1 - 2\phi\right)}{\epsilon} = - \kappa \phi(1-\phi).
	\label{eqn:rhs_eqmb}
\end{equation}
Now multiplying by $\hat{\sigma}$ and using the definition of $\mu$ (\cref{eqn:ac_mu}),
\begin{equation}
	-\mu =  \hat{\sigma}\epsilon\pdd{\phi}{x_j} - \hat{\sigma}\frac{\phi\left(1 - \phi\right)\left(1 - 2\phi\right)}{\epsilon} = - \hat{\sigma}\kappa \phi(1-\phi).
\end{equation}
Now using~\cref{eqn:rhs_eqmb} and multiplying by $-\pd{\phi}{x_i}$ we get, 
\begin{equation}
	\mu \pd{\phi}{x_i} =  \hat{\sigma}\kappa \phi \left(1 - \phi\right)\pd{\phi}{x_i} = \hat{\sigma}\kappa \epsilon \left(\pd{\phi}{x_k}\pd{\phi}{x_k}\right)^{1/2} \pd{\phi}{x_i}.
	\label{eqn:equivalence}
\end{equation}
It is important to note that in the right hand side of~\cref{eqn:equivalence}, the factor of $\phi(1-\phi)=\epsilon \left(\pd{\phi}{x_k}\pd{\phi}{x_k}\right)^{1/2}$ modifies the popular CSF model ($\hat{\sigma}\kappa\partial\phi/\partial x_i$) with a regularization kernel which is localized near the interface. 
\textcolor{ForestGreen}{We call $\hat{\sigma}\kappa \phi \left(1 - \phi\right)\pd{\phi}{x_i}$, the \textbf{\textit{localized CSF}} (LCSF) model.
The numerical results below confirm that this localization improves the accuracy of the surface tension model compared to the original CSF model.}\todo[color=ForestGreen]{R2: \#2.6}
\section{Results and discussion}
\label{sec:results_discussion}
In this section, using canonical tests, we numerically test the energy-based surface tension model by comparing its performance against CSF and LCSF. We use second-order finite differences on a staggered Cartesian grid in space and RK4 time-integration, as described in ~\citet{Mirjalili2021}.
\subsection{Spurious currents}
\label{subsec:spur_cur}
We consider a standard 2D benchmark for testing spurious currents generated by two-phase flow solvers~\citep{Williams1998, Popinet1999, Francois2006, Herrmann2008}. In a $1\times1$ domain with free-slip boundary conditions, at $t=0$, a circular drop with diameter $D=0.4$ is placed in the center of the domain with $\vec{u}=0$ everywhere. \textcolor{cyan}{Note that all numbers and parameters reported in this work are non-dimensional}\todo[color=cyan]{R2: \#1.3}. The density and viscosity of the two phases are equal, $\rho_1=\rho_2=300$ and $\mu_1=\mu_2=0.1$, the surface tension coefficient is $\sigma=1.0$, and the dimensionless parameter characterizing this problem is the Laplace number, $La=\sigma \rho D / { \mu }^{ 2 }=12000$. Any sustained flow in the numerical solution is due to errors in computing the surface tension force. 

\textcolor{ForestGreen}{To understand how the spurious currents evolve, in~\cref{fig:spurious_currents_time} we examine $Ca_{\infty}=({\norm{\vec{u}}_\infty\mu})/{\sigma}$ as a function of non dimensional time $t^*=\sigma t/( \mu D )$ using the CDI phase field model for all three (CSF, LCSF, and energy-based) surface tension formulations. We observe that after an initial transience, the energy-based model shows consistently lower spurious currents than the CSF and LCSF models. The transient higher values of spurious currents for the EB model at small times can be attributed to the interface being out of equilibrium. Based on these observations, and similar to ~\cite{Herrmann2008,Mirjalili_comparison}, we choose to compare the spurious currents at $t^*=250$. Evidently, this is a sufficiently large time for transient effects to have passed and for spurious currents to be fully developed for all three surface tension models due to the balance of surface tension and viscous forces.}\todo[color=ForestGreen]{R2: \#2.7}




Panel (a) of ~\cref{fig:spurious_currents} shows the convergence of $Ca_{\infty}$ at $t^*=250$ as we increase the number of mesh points ($N\times N$) for a fixed interfacial thickness, $\epsilon=1/32$. By successively refining the mesh, this analysis illustrates the property of the models in the continuous limit, thereby teasing out the PDE errors. It is clear that for all three chosen phase field equations, the energy-based model (denoted with EB) outperforms CSF and LCSF due to a better convergence order (second order for EB). \textcolor{ForestGreen}{It is worth noting that LCSF consistently yields lower spurious currents than CSF.} \todo[color=ForestGreen]{R2: \#2.6}

Panel (b) of ~\cref{fig:spurious_currents} shows the convergence of $Ca_{\infty}$ at $t^*=250$ as we approach the sharp-interface limit. An important property of a physical model is to converge to true sharp interface physics as $\epsilon\rightarrow 0$. As we reduce $\epsilon$, we refine the mesh via ${\Delta x}={32}^{1/2}{\epsilon}^{3/2}$ ~\citep{Jacqmin1999, Mirjalili_comparison}. In practice, to allow an integer number of mesh points, we successively refine the mesh by a factor of two while computing $\epsilon$ proportional to $\Delta x^{2/3}$. We observe that for all three phase field equations, the energy-based model converges faster (second order) to the sharp interface limit compared to the CSF and LCSF models. A faster convergence to the sharp interface limit has a practical benefit in terms of computational resource requirements, as it allows for the choice of larger $\epsilon$ values that usually require coarser spatio-temporal resolutions. \textcolor{ForestGreen}{ LCSF seems to consistently result in lower spurious currents than CSF for these simulations as well, although to a lesser degree than the fixed $\epsilon$ results shown in panel (a).}\todo[color=ForestGreen]{R2: \#2.6}


\begin{figure}[H]
	\centering
	\begin{tikzpicture}
		\begin{axis}[width=0.8\linewidth, height=0.4\linewidth,
            scaled y ticks=true,
			xlabel={$t^*$},
			ylabel={$Ca_{\infty} \left(t^*\right)$},
			legend entries={
                CDI with CSF,
                CDI with LCSF,
                CDI with EB
			},
			legend style={nodes={scale=0.65, transform shape}}, 
			xmin=0.0,xmax=400,
			scaled x ticks=true,
			cycle list/Set1,
			cycle multiindex* list={
				mark list*\nextlist
				Set1\nextlist
			},
			]
			\addplot [line width=1pt, index of colormap=0 of Set1] table [x index=0, y index = 1, 
            col sep=comma] {data/spurious_currents/Amplitude_64_CSF.csv};
            \addplot [line width=1pt, index of colormap=1 of Set1] table [x index=0, y index = 1, 
            col sep=comma] {data/spurious_currents/Amplitude_64_rCSF.csv};
            \addplot [line width=1pt, index of colormap=2 of Set1] table [x index=0, y index = 1, 
            col sep=comma] {data/spurious_currents/Amplitude_64_EBCSF.csv};
		\end{axis}
	\end{tikzpicture}
    \caption{\textit{\Cref{subsec:spur_cur} Spurious currents:} Variation of $Ca_{\infty}$ as a function of time for the CDI phase field model along with CSF (red line), LCSF (cyan line), and energy-based surface tension model (denoted by EB and green line). 
    \todo[color=ForestGreen,inline]{R2: \#2.6,\#2.7}}
    \label{fig:spurious_currents_time}
\end{figure}
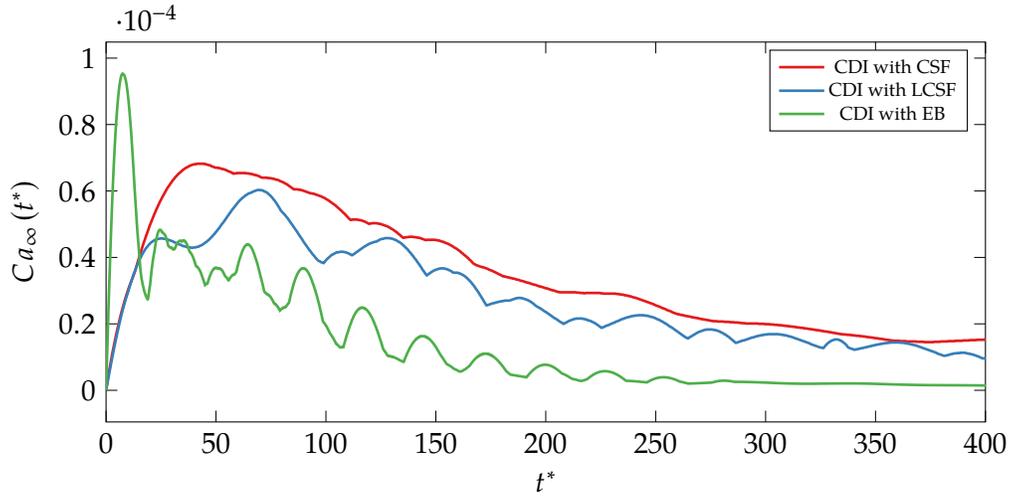
\begin{figure}[H]
	\centering
	\begin{tikzpicture}
		\begin{loglogaxis}[width=0.4\linewidth, scaled y ticks=true,
			xlabel={$\Delta x = 1/N$},
			ylabel={$Ca_{\infty}$},
			legend entries={
                CDI with EB, 
				CDI with CSF,
                CDI with LCSF,
				ISODI with EB, 
				ISODI with CSF,
                ISODI with LCSF,
				SB~\citep{Sun2007} with EB,
				SB~\citep{Sun2007} with CSF,
                SB~\citep{Sun2007} with LCSF
			},
			legend style={at={(0.5,-0.25)},anchor=north, nodes={scale=0.65, transform shape}}, 
			legend columns=3,
			title={\footnotesize(a) mesh convergence for fixed $\epsilon$ },
			xtick = {0.00390625, 0.0078125, 0.015625, 0.03125, 0.0625},
			xticklabel={
				\pgfkeys{/pgf/fpu=true}
				\pgfmathparse{exp(\tick)}%
				\pgfmathprintnumber[fixed relative, precision=2]{\pgfmathresult}
				\pgfkeys{/pgf/fpu=false}
			},
			scaled x ticks=true,
			cycle list/Set1,
			cycle multiindex* list={
				mark list*\nextlist
				Set1\nextlist
			},
			]
			\addplot [mark=*, mark options={solid}, line width=1pt, index of colormap=0 of Set1] table [x={h},y={cdi_eb},col sep=comma] {data/spurious_currents/sp_cur_fixed_epsilon_rcsf_added.csv};
			\addplot [mark=*, mark options={solid}, line width=1pt, index of colormap=0 of Set1, densely dashed] table [x={h},y={cdi_csf},col sep=comma] {data/spurious_currents/sp_cur_fixed_epsilon_rcsf_added.csv};
            \addplot [mark=*, mark options={solid}, line width=1pt, index of colormap=0 of Set1, dash dot] table [x={h},y={cdi_rcsf},col sep=comma] {data/spurious_currents/sp_cur_fixed_epsilon_rcsf_added.csv};
			\addplot [mark=square*, mark options={solid}, line width=1pt, index of colormap=1 of Set1] table [x={h},y={isodi_eb},col sep=comma] {data/spurious_currents/sp_cur_fixed_epsilon_rcsf_added.csv};
			\addplot [mark=square*, mark options={solid}, line width=1pt, index of colormap=1 of Set1, densely dashed] table [x={h},y={isodi_csf},col sep=comma] {data/spurious_currents/sp_cur_fixed_epsilon_rcsf_added.csv};
            \addplot [mark=square*, mark options={solid}, line width=1pt, index of colormap=1 of Set1, dash dot] table [x={h},y={isodi_rcsf},col sep=comma] {data/spurious_currents/sp_cur_fixed_epsilon_rcsf_added.csv};
			\addplot [mark=triangle*, mark options={solid}, line width=0.8pt, index of colormap=2 of Set1] table [x={h},y={sb_eb},col sep=comma] {data/spurious_currents/sp_cur_fixed_epsilon_rcsf_added.csv};
			\addplot [mark=triangle*, mark options={solid}, line width=0.8pt,  index of colormap=2 of Set1, densely dashed] table [x={h},y={sb_csf},col sep=comma] {data/spurious_currents/sp_cur_fixed_epsilon_rcsf_added.csv};
            \addplot [mark=triangle*, mark options={solid}, line width=0.8pt,  index of colormap=2 of Set1, dash dot] table [x={h},y={sb_rcsf},col sep=comma] {data/spurious_currents/sp_cur_fixed_epsilon_rcsf_added.csv};
			\addplot +[mark=none, black, line width=1pt, black, densely dotted] [domain=0.0078:0.065]{0.0065*x^1};
			\addplot +[mark=none, black, line width=1pt, blue, densely dotted] [domain=0.0078:0.065]{0.01*x^2};
			\logLogSlopeTriangle{0.28}{0.2}{0.07}{2}{blue};
			\logLogSlopeTriangle{0.28}{0.2}{0.62}{1}{black};
		\end{loglogaxis}
	\end{tikzpicture}
	\begin{tikzpicture}
		\begin{loglogaxis}[width=0.4\linewidth, scaled y ticks=true,
			xlabel={$\epsilon$},
			ylabel={$Ca_{\infty}$},
			legend entries={
                CDI with EB, 
				CDI with CSF,
                CDI with LCSF,
				ISODI with EB, 
				ISODI with CSF,
                ISODI with LCSF,
				SB~\citep{Sun2007} with EB,
				SB~\citep{Sun2007} with CSF,
                SB~\citep{Sun2007} with LCSF
			},
			legend style={at={(0.5,-0.25)},anchor=north, nodes={scale=0.65, transform shape}}, 
			legend columns=3,
			title={\footnotesize(b) $\epsilon$ convergence},
			xtick = {0.012401571, 0.019686266, 0.03125, 0.049606283},
			xticklabel={
				\pgfkeys{/pgf/fpu=true}
				\pgfmathparse{exp(\tick)}%
				\pgfmathprintnumber[fixed relative, precision=2]{\pgfmathresult}
				\pgfkeys{/pgf/fpu=false}
			},
			scaled x ticks=true,
			cycle list/Set1,
			cycle multiindex* list={
				mark list*\nextlist
				Set1\nextlist
			}
			]
			\addplot [mark=*, mark options={solid}, line width=1pt, index of colormap=0 of Set1] table [x={epsilon},y={cdi_eb},col sep=comma] {data/spurious_currents/sp_cur_varying_epsilon_rcsf_added.csv};
			\addplot [mark=*, mark options={solid}, line width=1pt, index of colormap=0 of Set1, densely dashed] table [x={epsilon},y={cdi_csf},col sep=comma] {data/spurious_currents/sp_cur_varying_epsilon_rcsf_added.csv};
            \addplot [mark=*, mark options={solid}, line width=1pt, index of colormap=0 of Set1, dash dot] table [x={epsilon},y={cdi_rcsf},col sep=comma] {data/spurious_currents/sp_cur_varying_epsilon_rcsf_added.csv};
			\addplot [mark=square*, mark options={solid}, line width=1pt, index of colormap=1 of Set1] table [x={epsilon},y={isodi_eb},col sep=comma] {data/spurious_currents/sp_cur_varying_epsilon_rcsf_added.csv};
			\addplot [mark=square*, mark options={solid}, line width=1pt, index of colormap=1 of Set1, densely dashed] table [x={epsilon},y={isodi_csf},col sep=comma] {data/spurious_currents/sp_cur_varying_epsilon_rcsf_added.csv};
            \addplot [mark=square*, mark options={solid}, line width=1pt, index of colormap=1 of Set1, dash dot] table [x={epsilon},y={isodi_rcsf},col sep=comma] {data/spurious_currents/sp_cur_varying_epsilon_rcsf_added.csv};
			\addplot [mark=triangle*, mark options={solid}, line width=0.8pt, index of colormap=2 of Set1] table [x={epsilon},y={sb_eb},col sep=comma] {data/spurious_currents/sp_cur_varying_epsilon_rcsf_added.csv};
			\addplot [mark=triangle*, mark options={solid}, line width=0.8pt,  index of colormap=2 of Set1, densely dashed] table [x={epsilon},y={sb_csf},col sep=comma] {data/spurious_currents/sp_cur_varying_epsilon_rcsf_added.csv};
            \addplot [mark=triangle*, mark options={solid}, line width=0.8pt,  index of colormap=2 of Set1, dash dot] table [x={epsilon},y={sb_rcsf},col sep=comma] 
            {data/spurious_currents/sp_cur_varying_epsilon_rcsf_added.csv};
			\addplot +[mark=none, black, line width=1pt, black, densely dotted] [domain=0.012:0.05]{0.0065*x^1};
			\addplot +[mark=none, black, line width=1pt, blue, densely dotted] [domain=0.012:0.05]{0.01*x^2};
			\logLogSlopeTriangle{0.28}{0.2}{0.07}{2}{blue};
			\logLogSlopeTriangle{0.25}{0.15}{0.68}{1}{black};
		\end{loglogaxis}
	\end{tikzpicture}	
	\caption{\textit{\Cref{subsec:spur_cur} Spurious currents:} Comparison of convergence of spurious currents generated by the proposed energy-based surface tension model (denoted by EB in the legend) against the CSF and LCSF models. Panel (a) shows mesh convergence of $Ca_{\infty}$ for a fixed $\epsilon$; Panel (b) shows the convergence of $Ca_{\infty}$ as $\epsilon$ approaches the sharp interface limit. Solid lines denote results for the proposed energy-based surface tension, dashed lines denote results for the CSF model, and dash-dotted lines denote the LCSF model. The black dotted line denotes a slope of 1, and the blue dotted line denotes a slope of 2.
    \todo[color=ForestGreen,inline]{R2: \#2.6}
    } 
	\label{fig:spurious_currents}
\end{figure}
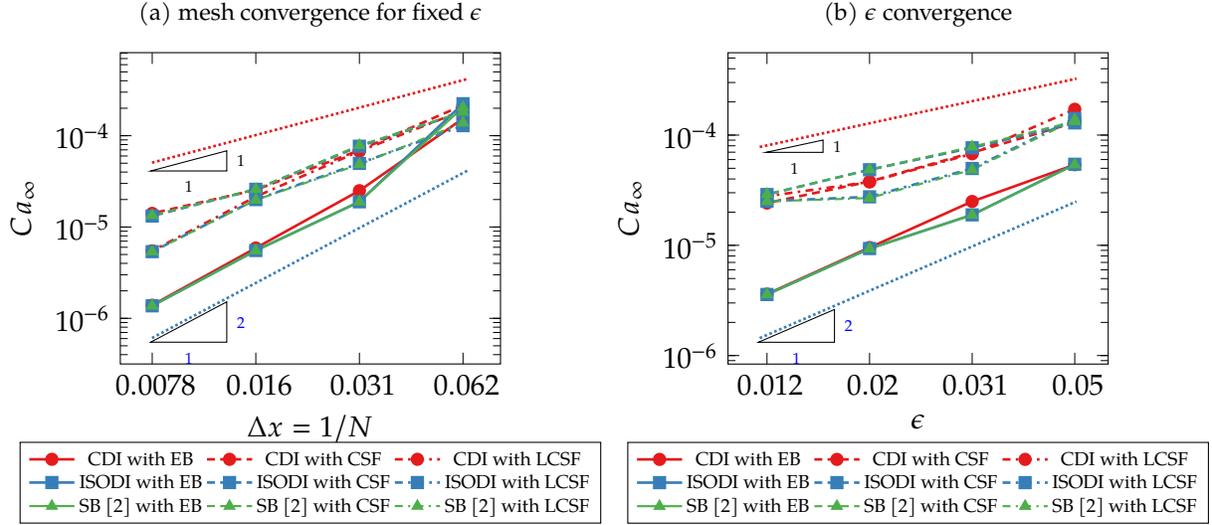

\subsection{Standing Wave}
\label{subsec:standing_wave}
By examining surface oscillations of a standing wave, one can assess the accuracy of a two-phase flow solver in capturing interactions among inertial, capillary, and viscous forces~\citep{Gerlach2006, Herrmann2008, Mirjalili_comparison}. 
We compare the accuracy of the energy-based surface tension model against the CSF and LCSF models in capturing the amplitude of a 2D standing wave as a function of time. In a $2\pi\times 2\pi$ domain with periodic boundary conditions in the $x$ direction and free-slip on the top and bottom walls, a small-amplitude wave is placed between two immiscible phases at $t=0$. The initial wave height is $h_{\text{wave}}(x,t=0)=y-y_0+A_{0}\cos(x-\Delta x/2)$, where $y_0=\pi$ and $A_0=2\pi/100$. The wavelength is $\lambda=2\pi$. The surface tension is $\sigma=2$, density of the two phases are ${\rho}_{1}={\rho}_{2}=1$, kinematic viscosity of both phases are $\nu_1=\nu_2=0.064720863$, and the time step is $\Delta t=0.003$. We measure the amplitude of the standing wave at $x=\Delta x/2$ as a function of time. 

Panels (a), (b), and (c) of ~\cref{fig:standing_wave} depict the normalized surface wave amplitude ($A/\lambda$) as a function of time for simulations performed 
using CDI, ISODI, and SB phase field methods, respectively. 
For panels (a), (b), and (c) of ~\cref{fig:standing_wave} we fix the value of $\epsilon=1/32$ and vary the mesh resolution.  We compare the performance of the energy-based surface tension model against CSF and the exact solution~(see~\citet{Prosperetti1981}) for each resolution. Notice that for all three models (panels (a), (b), and (c)), the energy-based surface tension model (solid lines and denoted by EB) outperforms CSF (dashed lines and denoted by CSF) when compared to the exact solution (except for the lowest resolution simulations with $16\times16$ mesh points).  
Additionally, it is important to note that as we refine the mesh, both CSF and the energy-based surface tension model converge to a solution, which implies a minimization of the spatial discretization error. Therefore, for the highest resolution, we predominantly observe a PDE error caused by choice of a finite $\epsilon$.  If we then inspect the converged solutions (purple solid line for the energy-based surface tension model and purple dashed line for the CSF model), it is clear that the PDE-level error committed using the energy-based surface tension model is much smaller than that of the CSF model for all three phase field methods (panels (a), (b), and (c)).  Therefore, we can infer from the results of panels (a), (b), and (c) that for a fixed $\epsilon$, the energy-based surface tension model has a smaller PDE error compared to CSF.  


Panel (d) of  ~\cref{fig:standing_wave} illustrates the convergence of the $l^2$ error computed between the surface wave amplitude and the theoretical solution~\citep{Prosperetti1981} as we approach the sharp interface limit (i.e., $\epsilon \rightarrow 0$).  
While reducing $\epsilon$, we refine the mesh via ${\Delta x}={32}^{1/2}{\epsilon}^{3/2}$.
For all three phase field equations, we observe that the energy-based surface tension force model (symbols with solid lines) is more accurate than the CSF model (symbols with dashed lines) while maintaining its convergence order of one.

\textcolor{ForestGreen}{
For panels (a), (b), and (c) of ~\cref{fig:standing_wave_lcsf} we fix the value of $\epsilon=1/32$ while varying the mesh resolution to compare the performance of the energy-based surface tension model against LCSF and the exact solution~(see~\citet{Prosperetti1981}). Similar to the comparison in~\cref{fig:standing_wave}, for all three phase field equations, the energy-based model (solid lines and denoted by EB) outperforms LCSF (dashed lines and denoted by LCSF) when compared against the exact solution. Furthermore, as we refine the mesh, LCSF and the energy-based surface tension model converge to a solution as the spatial discretization errors converge towards zero. The converged solutions show that the energy-based surface tension model (purple solid line) is much closer to the exact solution (solid black line) compared to the LCSF model (purple dashed line). This implies that the PDE-level error committed by the energy-based surface tension model is much smaller than the LCSF model for all three phase field methods.
}\todo[color=ForestGreen]{R2: \#2.6}

\textcolor{ForestGreen}{
Similar to panel (d) of ~\cref{fig:standing_wave}, panel (d) of  ~\cref{fig:standing_wave_lcsf} compares the convergence of the energy-based model and LCSF as we approach the sharp interface limit (i.e. $\epsilon \rightarrow 0$). We observe that the performance of the energy-based model (symbols with solid lines) is very close to the LCSF model, except for the CDI formulation. 
To understand why we see higher errors for CDI for the energy-based model, let us revisit Panel (a); we observe from panel (a) that the energy-based surface tension model outperforms the LCSF model as the interface is progressively resolved with more mesh points, specifically for resolutions more than $64 \times 64$ which corresponds to $\epsilon/\Delta x\ge 2$. For panel (d), we have $\epsilon/\Delta x = \sqrt{1/\left(32\epsilon\right)}$. Therefore, it is only as we approach $\epsilon = 0.0124$ for which $\epsilon/\Delta x \approx 1.58$ that we begin to see the energy-based model accuracy catching up to LCSF. Indeed, at $\epsilon = 0.0124$, we observe that the LCSF model accuracy is flattening, whereas the energy-based model is still converging at the same first-order rate.
}\todo[color=ForestGreen]{R2: \#2.6}

\textcolor{ForestGreen}{
In ~\cref{fig:standing_wave_energy}, we probe into the energetics of the standing wave problem. Namely, using CDI as the phase field model with $\epsilon=1/32$ on a $64\times64$ mesh, the amplitude evolution for the three different surface tension models is shown in panel (a). The time evolution of the kinetic energy (KE), surface energy (SE), and total energy (TE) for these simulations is shown in panel (b). SE is the system free energy given by $\frac{\hat{\sigma}}{\epsilon}\psi + \hat{\sigma}\epsilon\left({\pd{\phi}{x_j}}\right)\left({\pd{\phi}{x_j}}\right)$. Physically, energy is periodically exchanged between KE and SE while being dissipated by viscous dissipation. It is clear that during this process, compared to the energy-based surface tension model, the CSF and LCSF models exhibit unphysically high amplitude oscillations in  both KE and SE modes, which is correspondingly manifested by larger wave amplitudes in panel (a). The period of the oscillations seems to be also more erroneous for these surface tension models compared to the energy-based model. Overall, from panel (a) in conjunction with panel (b), it is evident that the energy-based model is more accurate than the CSF and LCSF models in flows involving dynamic interactions among surface energy, kinetic energy, and viscous dissipation.}\todo[color=ForestGreen]{R2:\#2.8}
\input{standing_wave_result.tex}

\section{Summary and perspective}
In this work, we provided theoretical justification for using ~\cref{eqn:eb_st} as an energy-based surface tension force model in conjunction with second-order phase field models. We described the connection of the energy-based surface tension model with a localized CSF model(~\cref{eqn:equivalence}), introduced in ~\cref{sec:equivalence_regularized_CSF}. Through extensive testing on two canonical problems, we revealed the relative accuracy of the energy-based surface tension model compared to CSF and LCSF.  These two cases allowed us to assess the accuracy of the model in capturing static equilibrium solutions and dynamic interactions involving the interplay of surface tension forces with inertial and viscous forces. We observed from~\cref{fig:spurious_currents} that the energy-based model leads to a faster rate of decay for spurious currents (second order) with respect to mesh resolution and interfacial thickness $\epsilon$ (the sharp interface limit) as compared to the CSF and LCSF models. This is analogous to observations from the Cahn-Hilliard model ~\citep{Jacqmin1999} and is intuitively expected since the equilibrium solution of the second-order phase field models is also a hyperbolic tangent. Additionally, from~\cref{fig:standing_wave},\ref{fig:standing_wave_lcsf}, and \ref{fig:standing_wave_energy}, we observed that the energy-based surface tension model is more accurate compared to the CSF and LCSF models in dynamically active scenarios involving exchanges between various modes of energy. 
 Both of these properties are highly sought after for accurate two-phase flow simulations. While our tests were two-dimensional, our findings can be readily generalized to three-dimensional settings. With this short note, we propose the model presented in ~\cref{eqn:eb_st} as an alternative to the CSF model for surface tension forces in two-phase flow simulations that utilize second-order phase field equations.

\section{Acknowledgements}
\label{sec:ack}
We acknowledge financial support from the Office of Naval Research (Grant N00014-15-1-2523) and Palo Alto Research Center (Grant 249996). The computations in this note were performed on the Yellowstone cluster at the Stanford HPC Center, supported through awards from Intel, the National Science Foundation, DOD HPCMP, and the Office of Naval Research. 

\bibliography{mybibfile}

\begin{thebibliography}{22}
\expandafter\ifx\csname natexlab\endcsname\relax\def\natexlab#1{#1}\fi
\providecommand{\url}[1]{\texttt{#1}}
\providecommand{\href}[2]{#2}
\providecommand{\path}[1]{#1}
\providecommand{\DOIprefix}{doi:}
\providecommand{\ArXivprefix}{arXiv:}
\providecommand{\URLprefix}{URL: }
\providecommand{\Pubmedprefix}{pmid:}
\providecommand{\doi}[1]{\href{http://dx.doi.org/#1}{\path{#1}}}
\providecommand{\Pubmed}[1]{\href{pmid:#1}{\path{#1}}}
\providecommand{\bibinfo}[2]{#2}
\ifx\xfnm\relax \def\xfnm[#1]{\unskip,\space#1}\fi
\bibitem[{Chiu and Lin(2011)}]{Chiu_and_Lin}
\bibinfo{author}{P.-H. Chiu}, \bibinfo{author}{Y.-T. Lin},
\newblock \bibinfo{title}{A conservative phase field method for solving
  incompressible two-phase flows},
\newblock \bibinfo{journal}{Journal of Computational Physics}
  \bibinfo{volume}{230} (\bibinfo{year}{2011}) \bibinfo{pages}{185--204}.
\bibitem[{Sun and Beckermann(2007)}]{Sun2007}
\bibinfo{author}{Y.~Sun}, \bibinfo{author}{C.~Beckermann},
\newblock \bibinfo{title}{Sharp interface tracking using the phase-field
  equation},
\newblock \bibinfo{journal}{Journal of Computational Physics}
  \bibinfo{volume}{220} (\bibinfo{year}{2007}) \bibinfo{pages}{626--653}.
\bibitem[{Mirjalili et~al.(2020)Mirjalili, Ivey, and
  Mani}]{Mirjalili_boundedness}
\bibinfo{author}{S.~Mirjalili}, \bibinfo{author}{C.~B. Ivey},
  \bibinfo{author}{A.~Mani},
\newblock \bibinfo{title}{A conservative diffuse interface method for two-phase
  flows with provable boundedness properties},
\newblock \bibinfo{journal}{Journal of Computational Physics}
  \bibinfo{volume}{401} (\bibinfo{year}{2020}) \bibinfo{pages}{109006}.
\bibitem[{Huang et~al.(2020)Huang, Lin, and Ardekani}]{Huang2020}
\bibinfo{author}{Z.~Huang}, \bibinfo{author}{G.~Lin}, \bibinfo{author}{A.~M.
  Ardekani},
\newblock \bibinfo{title}{Consistent and conservative scheme for incompressible
  two-phase flows using the conservative allen-cahn model},
\newblock \bibinfo{journal}{Journal of Computational Physics}
  \bibinfo{volume}{420} (\bibinfo{year}{2020}) \bibinfo{pages}{109718}.
\bibitem[{Jacqmin(1999)}]{Jacqmin1999}
\bibinfo{author}{D.~Jacqmin},
\newblock \bibinfo{title}{Calculation of two-phase navier-stokes flows using
  phase-field modeling},
\newblock \bibinfo{journal}{Journal of Computational Physics}
  \bibinfo{volume}{155} (\bibinfo{year}{1999}) \bibinfo{pages}{96--127}.
\bibitem[{Abels et~al.(2012)Abels, Garcke, and Gr{\"u}n}]{Abels2012}
\bibinfo{author}{H.~Abels}, \bibinfo{author}{H.~Garcke},
  \bibinfo{author}{G.~Gr{\"u}n},
\newblock \bibinfo{title}{Thermodynamically consistent, frame indifferent
  diffuse interface models for incompressible two-phase flows with different
  densities},
\newblock \bibinfo{journal}{Mathematical Models and Methods in Applied
  Sciences} \bibinfo{volume}{22} (\bibinfo{year}{2012})
  \bibinfo{pages}{1150013}.
\bibitem[{Khanwale et~al.(2020)Khanwale, Lofquist, Sundar, Rossmanith, and
  Ganapathysubramanian}]{Khanwale2020}
\bibinfo{author}{M.~A. Khanwale}, \bibinfo{author}{A.~D. Lofquist},
  \bibinfo{author}{H.~Sundar}, \bibinfo{author}{J.~A. Rossmanith},
  \bibinfo{author}{B.~Ganapathysubramanian},
\newblock \bibinfo{title}{Simulating two-phase flows with thermodynamically
  consistent energy stable cahn-hilliard navier-stokes equations on parallel
  adaptive octree based meshes},
\newblock \bibinfo{journal}{Journal of Computational Physics}
  \bibinfo{volume}{419} (\bibinfo{year}{2020}) \bibinfo{pages}{109674}.
\bibitem[{Brackbill et~al.(1992)Brackbill, Kothe, and Zemach}]{Brackbill1992}
\bibinfo{author}{J.~U. Brackbill}, \bibinfo{author}{D.~B. Kothe},
  \bibinfo{author}{C.~Zemach},
\newblock \bibinfo{title}{A continuum method for modeling surface tension},
\newblock \bibinfo{journal}{Journal of Computational Physics}
  \bibinfo{volume}{100} (\bibinfo{year}{1992}) \bibinfo{pages}{335--354}.
\bibitem[{Mirjalili and Mani(2021)}]{Mirjalili2021}
\bibinfo{author}{S.~Mirjalili}, \bibinfo{author}{A.~Mani},
\newblock \bibinfo{title}{Consistent, energy-conserving momentum transport for
  simulations of two-phase flows using the phase field equations},
\newblock \bibinfo{journal}{Journal of Computational Physics}
  \bibinfo{volume}{426} (\bibinfo{year}{2021}) \bibinfo{pages}{109918}.
\bibitem[{Gurtin(1996)}]{Gurtin1996}
\bibinfo{author}{M.~E. Gurtin},
\newblock \bibinfo{title}{Generalized ginzburg-landau and cahn-hilliard
  equations based on a microforce balance},
\newblock \bibinfo{journal}{Physica D: Nonlinear Phenomena}
  \bibinfo{volume}{92} (\bibinfo{year}{1996}) \bibinfo{pages}{178--192}.
\bibitem[{Shen and Yang(2010)}]{Shen2010}
\bibinfo{author}{J.~Shen}, \bibinfo{author}{X.~Yang},
\newblock \bibinfo{title}{A phase-field model and its numerical approximation
  for two-phase incompressible flows with different densities and viscosities},
\newblock \bibinfo{journal}{SIAM Journal on Scientific Computing}
  \bibinfo{volume}{32} (\bibinfo{year}{2010}) \bibinfo{pages}{1159}.
\bibitem[{Guo et~al.(2017)Guo, Lin, Lowengrub, and Wise}]{Guo2017}
\bibinfo{author}{Z.~Guo}, \bibinfo{author}{P.~Lin},
  \bibinfo{author}{J.~Lowengrub}, \bibinfo{author}{S.~M. Wise},
\newblock \bibinfo{title}{Mass conservative and energy stable finite difference
  methods for the quasi-incompressible navier--stokes--cahn--hilliard system:
  Primitive variable and projection-type schemes},
\newblock \bibinfo{journal}{Computer Methods in Applied Mechanics and
  Engineering} \bibinfo{volume}{326} (\bibinfo{year}{2017})
  \bibinfo{pages}{144--174}.
\bibitem[{Popinet(2018)}]{Popinet2018}
\bibinfo{author}{S.~Popinet},
\newblock \bibinfo{title}{Numerical models of surface tension},
\newblock \bibinfo{journal}{Annual Review of Fluid Mechanics}
  \bibinfo{volume}{50} (\bibinfo{year}{2018}) \bibinfo{pages}{49--75}.
\bibitem[{Kim(2005)}]{Kim2005}
\bibinfo{author}{J.~Kim},
\newblock \bibinfo{title}{A continuous surface tension force formulation for
  diffuse-interface models},
\newblock \bibinfo{journal}{Journal of Computational Physics}
  \bibinfo{volume}{204} (\bibinfo{year}{2005}) \bibinfo{pages}{784--804}.
\bibitem[{Abadi et~al.(2018)Abadi, Rahimian, and Fakhari}]{Haghani2018}
\bibinfo{author}{R.~H.~H. Abadi}, \bibinfo{author}{M.~H. Rahimian},
  \bibinfo{author}{A.~Fakhari},
\newblock \bibinfo{title}{Conservative phase-field lattice-boltzmann model for
  ternary fluids},
\newblock \bibinfo{journal}{Journal of Computational Physics}
  \bibinfo{volume}{374} (\bibinfo{year}{2018}) \bibinfo{pages}{668--691}.
\bibitem[{Williams et~al.(1998)Williams, Kothe, and Puckett}]{Williams1998}
\bibinfo{author}{M.~Williams}, \bibinfo{author}{D.~Kothe},
  \bibinfo{author}{E.~Puckett},
\newblock \bibinfo{title}{Accuracy and convergence of continuum surface tension
  models},
\newblock \bibinfo{journal}{Fluid Dynamics at Interfaces, Cambridge University
  Press, Cambridge}  (\bibinfo{year}{1998}) \bibinfo{pages}{294--305}.
\bibitem[{Popinet and Zaleski(1999)}]{Popinet1999}
\bibinfo{author}{S.~Popinet}, \bibinfo{author}{S.~Zaleski},
\newblock \bibinfo{title}{A front-tracking algorithm for accurate
  representation of surface tension},
\newblock \bibinfo{journal}{International Journal for Numerical Methods in
  Fluids} \bibinfo{volume}{30} (\bibinfo{year}{1999})
  \bibinfo{pages}{775--793}.
\bibitem[{Fran\c{c}ois et~al.(2006)Fran\c{c}ois, Cummins, Dendy, Kothe,
  Sicilian, and Williams}]{Francois2006}
\bibinfo{author}{M.~M. Fran\c{c}ois}, \bibinfo{author}{S.~J. Cummins},
  \bibinfo{author}{E.~D. Dendy}, \bibinfo{author}{D.~B. Kothe},
  \bibinfo{author}{J.~M. Sicilian}, \bibinfo{author}{M.~W. Williams},
\newblock \bibinfo{title}{{A balanced-force algorithm for continuous and sharp
  interfacial surface tension models within a volume tracking framework}},
\newblock \bibinfo{journal}{Journal of Computational Physics}
  \bibinfo{volume}{213} (\bibinfo{year}{2006}) \bibinfo{pages}{141--173}.
\bibitem[{Herrmann(2008)}]{Herrmann2008}
\bibinfo{author}{M.~Herrmann},
\newblock \bibinfo{title}{A balanced force refined level set grid method for
  two-phase flows on unstructured flow solver grids},
\newblock \bibinfo{journal}{Journal of Computational Physics}
  \bibinfo{volume}{227} (\bibinfo{year}{2008}) \bibinfo{pages}{2674--2706}.
\bibitem[{Mirjalili et~al.(2019)Mirjalili, Ivey, and
  Mani}]{Mirjalili_comparison}
\bibinfo{author}{S.~Mirjalili}, \bibinfo{author}{C.~B. Ivey},
  \bibinfo{author}{A.~Mani},
\newblock \bibinfo{title}{Comparison between the diffuse interface and volume
  of fluid methods for simulating two-phase flows},
\newblock \bibinfo{journal}{International Journal of Multiphase Flow}
  \bibinfo{volume}{116} (\bibinfo{year}{2019}) \bibinfo{pages}{221--238}.
\bibitem[{Gerlach et~al.(2006)Gerlach, Tomar, Biswas, and Durst}]{Gerlach2006}
\bibinfo{author}{D.~Gerlach}, \bibinfo{author}{G.~Tomar},
  \bibinfo{author}{G.~Biswas}, \bibinfo{author}{F.~Durst},
\newblock \bibinfo{title}{Comparison of volume-of-fluid methods for surface
  tension-dominant two-phase flows},
\newblock \bibinfo{journal}{International Journal of Heat and Mass Transfer}
  \bibinfo{volume}{49} (\bibinfo{year}{2006}) \bibinfo{pages}{740--754}.
\bibitem[{Prosperetti(1981)}]{Prosperetti1981}
\bibinfo{author}{A.~Prosperetti},
\newblock \bibinfo{title}{Motion of two superposed viscous fluids},
\newblock \bibinfo{journal}{Phys. Fluids} \bibinfo{volume}{24}
  (\bibinfo{year}{1981}) \bibinfo{pages}{1217--1223}.

\end{thebibliography}



\end{document}